\documentclass{elsart}

\usepackage{natbib}

\usepackage{amssymb}

\begin{document}

\begin{frontmatter}



\title{Quasi-stationary states and  incomplete violent relaxation in systems with long-range interactions}


\author{Pierre-Henri Chavanis}

\address{Laboratoire de Physique Th\'eorique, Universit\'e
Paul Sabatier\\ 118 route de Narbonne, 31062 Toulouse Cedex 4,
France}

\begin{abstract}
We discuss the nature of quasi-stationary states (QSS) with
non-Boltzmannian distribution in systems with long-range interactions
in relation with a process of incomplete violent relaxation based on
the Vlasov equation. We discuss several attempts to characterize these
QSS. We show that their distribution is {\it non-universal} and explain why
their prediction is difficult in general.
\end{abstract}

\begin{keyword}
Vlasov equation \sep long-range interactions \sep quasi-stationary states


\end{keyword}

\end{frontmatter}


\section{Quasi-stationary states: a generalized thermodynamics?}
\label{sec_qss}

It has been observed in many domains of physics [1] that Hamiltonian
systems with long-range interactions spontaneously organize into
coherent structures which persist for a long time. Some examples are
galaxies in astrophysics, jets and vortices in 2D geophysical flows, clusters in the HMF model etc. These  quasi-stationary
states (QSS) are usually {\it not} described by the Boltzmann
distribution. To account for this striking observational fact, some
authors have proposed to replace the Boltzmann entropy by the Tsallis
entropy
\begin{eqnarray}
\label{p1}
S_{q}[{f}]=-{1\over q-1}\int ({f}^{q}-{f}) d{\bf r}d{\bf v}.
\end{eqnarray}
The reason advocated is that the system is non-extensive so that
standard thermodynamics may not be applicable [2]. The maximization of
the Tsallis entropy at fixed mass and energy leads to
$q$-distributions of the form ${f}({\bf r},{\bf v})=\lbrack
\mu-{\beta(q-1)}\epsilon/q\rbrack^{1\over q-1}$, where
$\epsilon=v^{2}/2+\Phi({\bf r})$ is the individual energy and $\mu$,
$\beta$ are Lagrange multipliers. There are situations where these
distributions provide a good {\it fit} [2,3] of the QSS. However, there
exists other situations [4] that are described neither by the Boltzmann
nor by the Tsallis distribution [5]. We want to show that the
prediction of the QSS is relatively complicated and explain why. To
that purpose, we use classical methods of kinetic theory relying on
the Vlasov equation [1].

\section{Kinetic theory: the importance of the Vlasov equation}
\label{sec_vlasov}

To understand the physics of the problem, we first have to develop a
kinetic theory of systems with long-range interactions [1,6]. Such kinetic
theories have been developed for stellar systems, two-dimensional
vortices, the HMF model etc. They usually lead to a kinetic equation
for the distribution function $f({\bf r},{\bf v},t)$ of the form
\begin{equation}
{df\over dt}\equiv {\partial f\over\partial t}+{\bf v}\cdot {\partial f\over\partial {\bf r}}-\nabla\Phi\cdot {\partial f\over\partial {\bf v}}={1\over N^{\delta}}Q(f).
\label{p3}
\end{equation}
The left hand side, called the Vlasov term, describes an advection in
phase space due to the mean-field potential $\Phi({\bf r},t)=\int
u(|{\bf r}-{\bf r}'|) f({\bf r}',{\bf v}',t)d{\bf v}'d{\bf r}'$ where
$u$ is a binary potential of interaction. The right hand side takes
into account the effect of ``collisions'' (more generally
correlations) between particles and depends on the number $N$ of
particles. For long-range interactions, it usually scales like
$N^{-\delta}$ with $\delta\ge 1$ (in general $\delta=1$ but for 1D
systems $\delta>1$). This term, due to finite $N$ effects
(graininess), is responsible for the {\it collisional relaxation} of
the system towards ordinary statistical equilibrium. Indeed, in
general, the collision term cancels out only for the Boltzmann
distribution: $Q(f_{e})=0
\leftrightarrow f_{e}=Ae^{-\beta m\epsilon}$. However, due to its dependence with the number of particles,
the collisional relaxation time is of order $t_{R}\sim
N^{\delta}t_{D}$ where $t_{D}$ is the dynamical time. In general, this
timescale is huge and does not represent the regime of most physical
interest. In particular, the QSS mentioned previously form on a
timescale $t\sim t_{D}\ll t_{R}$ for which the collision term can be
neglected in a first approximation. In that case, the evolution is
collisionless and described by the Vlasov equation: $df/dt=0$.
Now, it has been understood, first in astrophysics by H\'enon, King
and Lynden-Bell in the 1960's, that the Vlasov equation, when coupled
to a long-range force like the gravitational force (Vlasov-Poisson
system) can undergo a form of {\it collisionless relaxation} on a very
short timescale $\sim t_{D}$. This is called {\it violent relaxation}
[7]. This general process explains the ubiquity of long-lived QSS in
systems with long-range interactions.  The fine-grained DF $f({\bf
r},{\bf v},t)$ which is solution of the Vlasov equation never achieves
equilibrium but develops intermingled filaments at smaller and smaller
scales due to {\it phase mixing}. However, if we locally average over
these filaments, the coarse-grained DF $\overline{f}({\bf r},{\bf
v},t)$ is expected to converge toward a steady state which is a stable
stationary solution of the Vlasov equation. This is called weak
convergence in mathematics. The kinetic theory explains that the
lifetime of the QSS scales as a power $N^{\delta}$ of the number of
particles. The kinetic theory also explains that when the
$t\rightarrow +\infty$ limit is taken before the $N\rightarrow
+\infty$ limit we get the Boltzmann statistical equilibrium state
(collisional relaxation) but when the $N\rightarrow +\infty$ limit is
taken before the $t\rightarrow +\infty$ limit we get a QSS which is a
stable stationary solution of the Vlasov equation usually different
from the Boltzmann distribution (collisionless regime). This
non-commutation of the limits $N\rightarrow +\infty$ and $t\rightarrow
+\infty$ has been observed numerically by [2] although they did not
give an explanation in terms of the Vlasov equation, as was done in
[1] (see Sec. 6.5) but rather in terms of Tsallis generalized
thermodynamics. 
On short timescales $\sim t_{D}$, the system undergoes violent
relaxation and reaches a stationary solution of the Vlasov equation
(on the coarse-grained scale). Since the Vlasov equation admits an
infinite number of stationary solutions, the solution (QSS)
effectively selected by the evolution is difficult to predict (see
below). On intermediate timescales $t_{D}<t<t_{R}$, the system passes
by a succession of quasi-stationary states that are quasi-stationary
solutions of the Vlasov equation $f(\epsilon,t)$ slowly evolving with
time due to ``collisions'' (finite $N$ effects). In astrophysics, this
slow collisional evolution is governed by the
orbit-averaged-Fokker-Planck equation. On a longer timescale $\sim
t_{R}$, collisions finally select the Maxwell distribution among all
stationary solutions of the Vlasov equation (the fate of gravitational
systems is peculiar due to the absence of strict statistical
equilibrium and the process of evaporation or collapse).  These
different regimes have been observed for different physical systems [1,6]
and are illustrated numerically by Yamaguchi {\it et al.} [8] for the HMF model.

\section{Lynden-Bell's theory of violent relaxation}
\label{sec_lb}

In a seminal paper, Lynden-Bell [7] tried to {\it predict} the QSS
resulting from violent relaxation assuming that the system mixes well
(ergodicity) and using arguments of statistical mechanics. However,
the statistical mechanics of the Vlasov equation is peculiar because
of the presence of an infinite set of constraints: the Casimirs
$I_{h}=\int h(f)d{\bf r}d{\bf v}$ (for any function $h$) which contain
all the moments $I_{n}=\int f^{n}d{\bf r}d{\bf v}$ of
the distribution function. These are sort of ``hidden constraints'' [5]
because they are not accessible from the (observed) coarse-grained DF
$\overline{f}$ since $\int
\overline{f^{n}}d{\bf r}d{\bf v}\neq \int
\overline{f}^{n}d{\bf r}d{\bf v}$ for $n\neq 1$. Therefore, the proper density
probability to consider in the statistical theory of violent
relaxation is $\rho({\bf r},{\bf v},\eta)$ which gives the density
probability of finding the level of DF $f=\eta$ in $({\bf r},{\bf v})$
in phase space. Note that we make the statistical mechanics of a {\it
field}, the distribution function, not the statistical mechanics of
{\it discrete particles}. From this density probability, we can
construct all the coarse-grained moments $\overline{f^{n}}=\int
\rho
\eta^{n}d\eta$ including the coarse-grained DF $\overline{f}=\int \rho
\eta d\eta$. We can now introduce a mixing entropy from a
combinatorial analysis [7,5] like in Boltzmann's traditional approach. In
the context of the Vlasov equation, this entropy is a functional of
$\rho({\bf r},{\bf v},\eta)$ of the form
\begin{equation}
S_{L.B.}[\rho]=-\int \rho({\bf r},{\bf v},\eta)\ln\rho({\bf r},{\bf v},\eta)d{\bf r}d{\bf v}d\eta.
\label{p5}
\end{equation}
The Lynden-Bell entropy (\ref{p5}) is the proper form of Boltzmann
entropy taking into account the specificities of the Vlasov
equation. Assuming ergodicity (efficient mixing), the QSS is obtained
by maximizing $S_{L.B.}[\rho]$ at fixed mass, energy and Casimir
invariants. This leads to an optimal $\rho_{*}({\bf r},{\bf
v},\eta)={1\over Z({\bf r})}\chi(\eta)e^{-\eta(\beta\epsilon+\alpha)}$ from
which we obtain the optimal coarse-grained distribution
\begin{equation}
\label{p6}
\overline{f}={\int \chi(\eta)\eta e^{-\eta(\beta\epsilon+\alpha)}d\eta\over\int \chi(\eta) e^{-\eta(\beta\epsilon+\alpha)}d\eta}=f_{L.B.}(\epsilon),
\end{equation}
where $\alpha$ and $\beta$ are the usual Lagrange multipliers
associated with the conservation of mass $M$ and energy $E$ while
$\chi(\eta)={\rm exp}\lbrace -\sum_{n}\alpha_{n}\eta^{n}\rbrace$ takes
into account the conservation of all the Casimirs invariants $I_{n}$
[5]. We note that, even if the system mixes well (ergodicity), the
coarse-grained DF predicted by Lynden-Bell may {\it differ} from the
Boltzmann distribution: $f_{L.B.}(\epsilon)\neq Ae^{-\beta\epsilon}$
({in general}) due to the presence of the Casimir invariants. In the
simplest case (two-levels approximation) the DF predicted by
Lynden-Bell is similar to the Fermi-Dirac statistics:
$f_{L.B.}=\eta_{0}/(1+e^{\beta\epsilon-\mu})$ [9]. More generally,
the coarse-grained DF (\ref{p6}) is a sort of {\it superstatistics}
[5,10] as it is expressed as a superposition of Boltzmann distributions
(universal) weighted by a non-universal factor $\chi(\eta)$ depending
on the initial conditions. Furthermore, like for the Beck-Cohen
superstatistics, the coarse-grained DF (\ref{p6}) maximizes a
``generalized entropy'' in $\overline{f}$-space $S[\overline{f}]=-\int
C(\overline{f})d{\bf r}d{\bf v}$ with
$C(\overline{f})=-\int^{\overline{f}} \lbrack (\ln
\hat\chi)'\rbrack^{-1}(-x)dx$ (where
$\hat{\chi}(\Phi)=\int_{0}^{+\infty}\chi(\eta)e^{-\eta\Phi}d\eta$) at
fixed mass and energy [5]. The distribution (\ref{p6}) and the
corresponding entropy $S[\overline{f}]$ are non-universal. However, a
general prediction of the Lynden-Bell statistical theory [7] is that
the QSS is a stationary solution of the Vlasov equation of the form
$\overline{f}=\overline{f}(\epsilon)$ with
$\overline{f}'(\epsilon)<0$: the DF depends only on the energy and is
monotonically decreasing. Furthermore, $\overline{f}({\bf r},{\bf
v})\le {\rm max}_{{\bf r},{\bf v}} f({\bf r},{\bf v},t=0)$: the
coarse-grained DF is bounded by the maximum value of the initial DF.

\section{Incomplete violent relaxation}
\label{sec_incomplete}

The Lynden-Bell DF (\ref{p6}) is the proper prediction of the QSS when
mixing is efficient (ergodic) during violent relaxation. There are
situations where the Lynden-Bell prediction works relatively well, e.g.
[11]. However, it has been recognized in many other occasions [4] that
mixing is {\it not} efficient enough to sustain the hypothesis of
ergodicity on which the theory is built so that the prediction of
Lynden-Bell fails in practice: $f_{QSS}\neq f_{LB}(\epsilon)$ ({in \
general}) [1]. This is particularly obvious in the case of
self-gravitating systems because the Lynden-Bell entropy has no
maximum at fixed (finite) mass and energy [7,9]. What can we do to account
for {\it incomplete relaxation}?

A first possibility would be to change the form of entropy. For
example, we could try to use the ``generalized thermodynamics'' of
Tsallis. However, as explained previously, the proper density probability is
$\rho({\bf r},{\bf v},\eta)$ so that the proper form of Tsallis
entropy in the context of Vlasov systems is [5]:
\begin{equation}
S_{q}[\rho]=-{1\over q-1}\int \left\lbrack \rho^{q}({\bf r},{\bf
v},\eta)-\rho({\bf r},{\bf v},\eta)\right\rbrack d{\bf r}d{\bf v}d\eta,
\label{p10}
\end{equation}
instead of (\ref{p1}). For $q=1$ it returns the Lynden-Bell entropy
(\ref{p5}).  In this line of thought, $q$ would measure the efficiency
of mixing ($q=1$ if the evolution is ergodic). For $q\neq 1$, the
functional (\ref{p10}) could describe non-ergodic behaviours
(incomplete mixing). However, it is not clear why all non-ergodic
behaviours could be described by a simple functional such as
(\ref{p10}). Tsallis entropy may describe
a certain type of non-ergodicity (fractal or multifractal) with
phase-space structures but probably not all of them. In particular,
observations of elliptical galaxies in astrophysics [4] do {\it not}
favour Tsallis form of entropy since elliptical galaxies are not {\it
stellar polytropes} that would be the prediction based on the
$q$-entropy [5]. Therefore, $f_{QSS}\neq f_{q}(\epsilon)$ ({in \
general}).

Another possibility is to keep the Lynden-Bell entropy as the most
fundamental entropy of the problem but develop a {\it dynamical}
theory of violent relaxation. Indeed, if relaxation is incomplete, we
must understand why. Qualitatively, the collisionless relaxation is
driven by the fluctuations of the field $\Phi({\bf r},t)$. Now, these
fluctuations can vanish before the system had time to relax completely
so that the system can remain frozen in a stationary state of the
Vlasov equation which is not the most mixed state. By using a
phenomenological Maximum Entropy Production Principle (MEPP), we have
proposed in [12] to describe the out-of-equilibrium evolution of the
probability density $\rho({\bf r},{\bf v},\eta,t)$ by a relaxation
equation of the form
\begin{equation}
{\partial\rho\over\partial t}+{\bf v}\cdot {\partial\rho\over\partial {\bf r}}-\nabla\Phi\cdot {\partial\rho\over\partial {\bf v}}={\partial\over\partial {\bf v}}\cdot \biggl\lbrace D({\bf r},{\bf v},t)\biggl\lbrack {\partial\rho\over\partial {\bf v}}+\beta(t)(\eta-\overline{f})\rho {\bf v}\biggr\rbrack\biggr\rbrace.
\label{p13}
\end{equation}
If the diffusion coefficient were constant, this equation would relax
towards the Lynden-Bell distribution. However, it is argued in [12]
that $D$ is {\it not} constant. Indeed, the relaxation is driven by
the fluctuations of the field $\Phi$, itself induced by the
fluctuations of $f$, so that the diffusion coefficient should be
proportional to $\overline{\tilde
f^{2}}=\overline{f^{2}}-\overline{f}^{2}$ and vanish in the regions of
phase space where these fluctuations vanish. Moreover, as the system
approaches (quasi)-equilibrium, the fluctuations of the field
$\delta\Phi$ are less and less efficient so that the diffusion
coefficient should also decay with time. For these reasons, it can
become very small $D({\bf r},{\bf v},t)\rightarrow 0$ in certain
regions of phase space (where mixing is not very efficient) and for
large times (as the fluctuations weaken). The diffusion coefficient
can also rapidly decay with the velocity. The vanishing of the
diffusion coefficient can ``freeze'' the system in a subdomain of
phase space (bubble) and account for incomplete relaxation and
non-ergodicity. The relaxation equation (\ref{p13}) should then tend
to a distribution which is only {\it partially mixed} and which is
usually different from the Lynden-Bell and the Tsallis distributions.
However, this approach demands to solve a dynamical equation
(\ref{p13}) -smoother than the Vlasov equation- to predict the
metaequilibrium state. The idea is that, in case of incomplete
relaxation (non-ergodicity), the prediction of the QSS is impossible
without considering the dynamics: it depends on the ``route to
equilibrium''.

Maybe, we have to accept that, in general, the QSS is {\it
unpredictable} in case of incomplete relaxation. We can expect,
however, that $\overline{f}_{QSS}({\bf r},{\bf v})$ is a stable
stationary solution of the Vlasov equation. We are thus led to
construct stable stationary solutions of the Vlasov equation is order
to {\it reproduce} observations. The Vlasov equation admits an
infinite number of stationary solutions given by the Jeans theorem,
but not all of them are stable. Of course, only stable solutions must
be considered and their selection is a difficult problem. If we
restrict ourselves to DF of the form $f=f(\epsilon)$ with
$f'(\epsilon)<0$ depending only on the energy (in astrophysics, such
distributions characterize a subclass of {\it spherical} stellar
systems), it is possible to provide a simple criterion of nonlinear
dynamical stability. Such DF extremize a functional of the form
$H[f]=-\int C(f)d{\bf r}d{\bf v}$, where $C$ is convex, at fixed mass
and energy [13]. Indeed, the first variations $\delta H-\beta\delta
E-\alpha\delta M=0$ yield a DF of the form $f=F(\beta\epsilon+\alpha)$
with $F(x)=(C')^{-1}(-x)$ montonically decreasing.  If, furthermore,
the DF {\it maximizes} this functional (at fixed $E$, $M$), then it is
nonlinearly dynamically stable with respect to the Vlasov equation
[13,8,14]. The intrinsic reason is that $H[f]$, $E[f]$ and $M[f]$ are
individually conserved by the Vlasov equation. Therefore, if
$f_{0}({\bf r},{\bf v})$ is the maximum of $H[f]$ (at fixed $E$, $M$),
a small perturbation $f({\bf r},{\bf v},t)$ will remain close (in some
norm) to this maximum.  It is important to note that this criterion of
nonlinear dynamical stability is remarkably consistent with the
phenomenology [5] of violent relaxation if we view the relevant DF as
the {\it coarse-grained} DF. Indeed, during mixing
$d\overline{f}/dt\neq 0$ and the functionals $H[\overline{f}]=-\int
C(\overline{f})d{\bf r}d{\bf v}$ calculated with the coarse-grained DF
increase ($-H$ decrease) in the sense that $H[\overline{f}({\bf
r},{\bf v},t)]\ge H[\overline{f}({\bf r},{\bf v},0)]$ for $t\ge 0$
where it is assumed that initially the system is not mixed:
$\overline{f}({\bf r},{\bf v},0)={f}({\bf r},{\bf v},0)$. Because of
this property similar to the Boltzmann $H$-theorem in kinetic theory
\footnote{Note that the Vlasov equation does not single out a unique
H-function contrary to the Boltzmann equation (the above inequality is
true for all $H$-functions) and the time evolution of the
$H$-functions is not necessarily monotonic (nothing is implied
concerning the relative values of $H(t)$ and $H(t')$ for $t,t'>0$).},
$H[\overline{f}]$ are called generalized $H$-functions [13]. By
constrast, $E[\overline{f}]$ and $M[\overline{f}]$ are approximately
conserved. Therefore, this generalized {\it selective decay principle}
[5] (decrease of $-H[\overline{f}]$ at fixed $E$, $M$) due to phase mixing and
coarse-graining can explain {\it how} $\overline{f}$ can possibly
reach a maximum of $H$ at fixed mass and energy (while $H[f]$ is
rigorously conserved on the fine-grained scale). After mixing
$d\overline{f}/dt= 0$ and the functionals $H[\overline{f}]$ (as well
as $E[\overline{f}]$ and $M[\overline{f}]$) are conserved by the
coarse-grained flow. Therefore, if $\overline{f}$ has reached (as a
result of mixing) a maximum of $H$, it will be nonlinearly dynamically
stable with respect to coarse-grained perturbations (after mixing) in
virtue of the above-mentioned stability result.

In reality, the problem is more complicated because the system can
converge toward a stationary solution of the Vlasov equation which
does not depend on the energy $\epsilon$ alone, and thus which does
not maximize an $H$-function at fixed $E$, $M$. For example, the
velocity distribution of stars in elliptical galaxies is anisotropic
and depends on the angular momentum $J=|{\bf r}\times {\bf v}|$ in
addition to energy $\epsilon$. Furthermore, real stellar systems are
in general not spherically symmetric so their DF does not only depend
on $\epsilon$ and $J$.  Therefore, more general stationary solutions
of the Vlasov equation must be constructed in consistency with the
Jeans theorem. Elliptical galaxies are well relaxed (in the sense of
Lynden-Bell) in their inner region (leading to an isotropic isothermal
core \footnote{One success of Lynden-Bell's theory of violent
relaxation is precisely to explain the isothermal cores of elliptical
galaxies without recourse to ``collisions'' whose effect manifests
itself on a much longer timescale.} with density profile $\sim
r^{-2}$) while they possess radially anisotropic envelopes (with
density profile $\sim r^{-4}$). Stiavelli \& Bertin [4] introduced an
$f^{(\infty)}$ model of the form
$f^{(\infty)}=A(-\epsilon)^{3/2}e^{-a\epsilon-cJ^{2}/2}$ for $\epsilon\le 0$ (and $f=0$ otherwise) based on the
possibility that the {\it a priori} probabilities of microstates are
not equal due to kinetic constraints. This model reproduces many
properties of ellipticals but it has the undesired feature of being
``too isotropic''. Then, they introduced another model based on a
modification of the Lynden-Bell statistical theory. They considered
the maximization of the Boltzmann entropy (in Lynden-Bell's sense) at
fixed mass, energy {\it and} a third global quantity $Q=\int
J^{\nu}|\epsilon|^{-3\nu/4}f d{\bf r}d{\bf v}$ which is argued to be
approximately conserved during violent relaxation. This variational
principle results in a family of $f^{(\nu)}$ models $f^{(\nu)}=A\ {\rm
exp}\lbrack -a\epsilon-d\left ({J^{2}/ |\epsilon|^{3/2}}\right
)^{\nu/2}\rbrack$. These models are able to fit products of $N$-body
simulations over nine orders of magnitude in density and to reproduce
the de Vaucouleur's $R^{1/4}$ law (or more general $R^{1/n}$ laws) of
ellipticals.  The introduction of additional constraints in the
variational principle could be a way to take into account effects of
incomplete violent relaxation.

We note finally that the Vlasov equation can have a very complicated,
non-ergodic, dynamics.  For example, in the gravitational 1D
Vlasov-Poisson system, phase-space holes which block the relaxation
towards the Lynden-Bell distribution (incomplete relaxation) have been
observed [15]. In that case, the system does not even relax towards a
stationary state of the Vlasov equation but develops everlasting
oscillations. Rapisarda \& Pluchino [16] have also observed transient
phase-space structures in their $N$-body simulations of the HMF model
and they have proposed an interesting analogy with glassy dynamics.
These results are not necessarily in contradiction with the Vlasov
equation (as they beleive), even for the inhomogeneous situations that
they consider. It would be of interest to check whether these
phase-space structures can also be obtained by solving the Vlasov
equation for the HMF model.

\section{Summary and conclusion}
\label{sec_concl}

Non-Boltzmannian distributions appear in the study of Hamiltonian
systems with long-range interactions. For these systems, the
collisional relaxation time towards statistical equilibrium (Boltzmann
distribution) is huge because it increases as a power of the number of
particles $N$. Therefore, the evolution of the system is described by
the Vlasov equation on a very long timescale [1]. Now, because of
phase mixing and violent relaxation, the Vlasov equation can
spontaneously lead to the formation of coherent structures: galaxies
in astrophysics, jets and vortices in hydrodynamics, clusters in the
HMF model, bars in disk galaxies... These QSS are (nonlinearly)
dynamically stable stationary solutions of the Vlasov equation which
are not necessarily described by the Boltzmann distribution. Indeed,
the Vlasov equation admits an infinite number of stationary solutions
and the system can be trapped in one of them. In general, the QSS is
{\it non-universal} as it depends: (i) on the {\it detailed} structure
of the initial conditions (through the Casimirs, in addition to mass
and energy) and (ii) on the efficiency of mixing (ergodicity). In this
Vlasov context, the Tsallis distributions are {\it particular}
stationary solutions of the Vlasov equation which correspond to what
are called {\it stellar polytropes} in astrophysics [5,14]. In
addition, the Tsallis functional $S_{q}[\overline{f}]$ is a {\it
particular} $H$-function [13], not an entropy (which would be a
functional $S_{q}[\rho]$). Its maximization at fixed mass and energy
yields a criterion of nonlinear dynamical stability with respect to
the Vlasov equation [8,13,14], not a criterion of generalized
thermodynamical stability. A formal {\it thermodynamical analogy} [14]
can however be developed to investigate the nonlinear dynamical
stability problem.




\end{document}